\pdfoutput=1

\documentclass[11pt]{article}
\usepackage{authblk}
\usepackage[]{ACL2023}

\usepackage{times}
\usepackage{latexsym}
\usepackage{graphicx}
\usepackage{multirow}
\usepackage{multicol}
\usepackage{amsmath}
\usepackage{subfigure}
\usepackage{float}
\usepackage{soul}

\usepackage{algorithm}
\usepackage{algorithmic}
\usepackage{array}
\usepackage{booktabs}
\usepackage{amsmath}
\usepackage{multirow}
\usepackage{newfloat}
\usepackage{listings}
\floatstyle{ruled}
\newfloat{listing}{tb}{lst}{}
\floatname{listing}{Listing}

\usepackage{subcaption} 
\usepackage[T1]{fontenc}

\usepackage[utf8]{inputenc}

\usepackage{microtype}
\usepackage{amsmath}

\usepackage{inconsolata}

\title{Mistral-SPLADE: LLMs for better Learned Sparse Retrieval}

\author[*]{\bf Meet Doshi}
\author[$^\ddagger$]{\bf Vishwajeet Kumar}
\author[$^\ddagger$]{\bf Rudra Murthy}
\author[$^\ddagger$]{\bf Vignesh P}
\author[$^\ddagger$]{\bf Jaydeep Sen}

\affil[*]{Indian Institute of Technology Bombay, India}
\affil[$^\ddagger$]{IBM Research, India}
\affil[*]{\texttt{meetdoshi@cse.iitb.ac.in}}
\affil[$^\ddagger$]{\texttt{\{vishk024,jaydesen,rmurthyv\}@in.ibm.com}}

\begin{document}
\maketitle

\begin{abstract}
Learned Sparse Retrievers (LSR) have evolved into an effective retrieval strategy that can bridge the gap between traditional keyword-based sparse retrievers and embedding-based dense retrievers. At its core, learned sparse retrievers try to learn the most important semantic keyword expansions from a query and/or document which can facilitate better retrieval with overlapping keyword expansions. LSR like SPLADE has typically been using encoder only models with MLM (masked language modeling) style objective in conjunction with known ways of retrieval performance improvement such as hard negative mining, distillation, etc. In this work, we propose to use decoder-only model for learning semantic keyword expansion. We posit, decoder only models that have seen much higher magnitudes of data are better equipped to learn keyword expansions needed for improved retrieval. We use Mistral as the backbone to develop our Learned Sparse Retriever similar to SPLADE and train it on a subset of sentence-transformer data which is often used for training text embedding models. Our experiments support the hypothesis that a sparse retrieval model based on decoder only large language model (LLM) surpasses the performance of existing LSR systems, including SPLADE and all its variants. The LLM based model (Echo-Mistral-SPLADE) now stands as a state-of-the-art learned sparse retrieval model on the BEIR text retrieval benchmark.

\end{abstract}

\textbf{Disclaimer}

This work is still in progress, we aim to present preliminary results for learned sparse retrievers using LLMs.

\section{Introduction}

Information Retrieval (IR) has been greatly influenced by traditional systems like BM25, which use lexical matching and inverted indices for efficient retrieval. These methods have been the backbone of search engines for years. However, Pre-trained Language Models (PLMs) like BERT \cite{devlin-etal-2019-bert} have brought a major shift towards contextualized semantic matching. While bag-of-words (BOW) models remain strong baselines \cite{critical-neural-ir}, they suffer from the long-standing vocabulary mismatch problem, where relevant documents might not contain terms that appear in the query. Thus, there have been attempts to substitute standard Bag of Words approaches with neural rankers.

In modern IR, dense representations combined with Approximate Nearest Neighbors (ANN) search has become the standard for first-stage ranking to reduce inference and maintain semantic representation similarity. These models have improved their in-domain performance thanks to better training pipelines. Despite these improvements, the generalization ability of these models has been questioned, especially with the BEIR benchmark \cite{beir} showing that traditional systems like BM25 can outperform neural retrievers in various IR tasks and in some cases only show incremental gains. Moreover, dense models have also not been able to explicitly model term matching.

Recent advances in the ``lexical representation space'' through learned sparse retrieval (LSR) have shown encouraging results for fast sparse inference \cite{spladev2, splade, spladev3}. This approach aims to combine the strengths of lexical bag-of-words and vector embedding algorithms by learning sparse representations that can be integrated with inverted indexing techniques. These methods often involve term weighting and expansion, using the benefits of explicit lexical matching and the efficiency of inverted indices. LSR methods have shown good generalization capabilities, both in inference latency and IR performance.

Despite their potential, sparse retrieval models have not yet been fully used to their full potential. We investigate LSR systems using LLMs and demonstrate substantial performance improvements over existing SPLADE based models and its variants. Additionally, we examine the effects of LSR training on interpretability, offering insights into the future of learned sparse retrieval systems.

\begin{figure*}
    \centering
    \includegraphics[width=\linewidth]{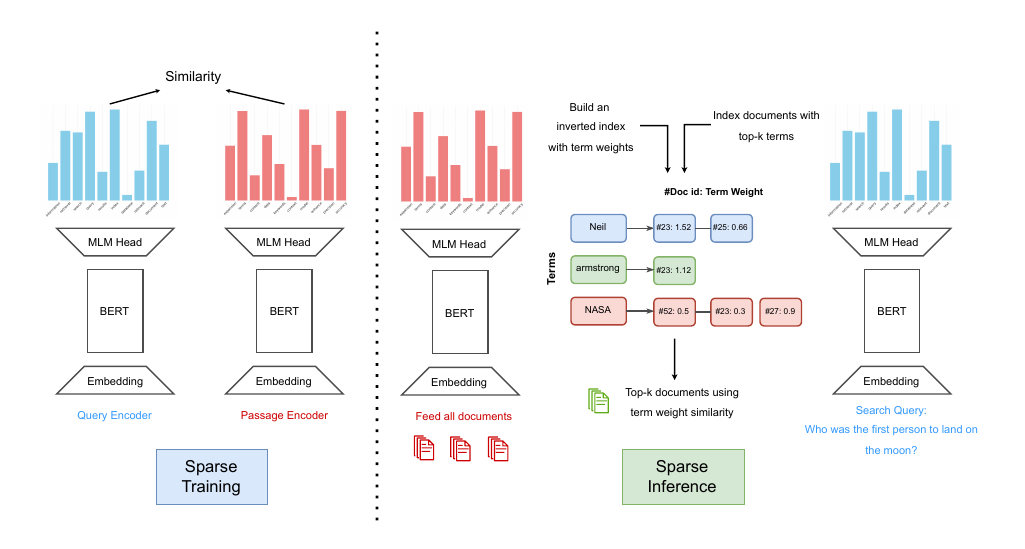}
    \caption{Learned Sparse Retrieval: Training vs Inference}
    \label{fig:enter-label}
\end{figure*}

\section{Related Work}

We review the literature on different retrieval paradigms including embedding based dense models with an Approximate Nearest Neighbor (ANN) search and more recent attempts to do learned sparse retrievals which aims to learn semantic expansion of contextual synonyms to do better retrieval and retain interpretability. We also touch upon the latest research advances that attempts to utilize causal only decoder models to learn better dense representations, which as we will see in our work, is also suitable for learned sparse retrievers. 

\subsection{Dense Representation Learning}

Dense retrieval has become a prevalent method for applying PLM retrievers in Information Retrieval (IR) where the models typically rely on the \texttt{[CLS]} token representation. Various training strategies have been proposed to enhance learned representations including but not limited to distillation \cite{distillation1, distillation2, distillation3, distillation4}, hard negative mining \cite{hnm1, hnm2, hnm3, hnm4}, pre-training \cite{pretraining1, pretraining2} and also all of them combined. Models such as DPR \cite{dpr} and ANCE \cite{hnm2} have been recently evaluated on generalization tasks, as demonstrated in the BEIR benchmark \cite{beir}. Colbert \cite{colbert} which is a dense representation model utilizes a late-interaction mechanism that allows fine-grained term-level matching at the expense of higher indexing cost and latency.

\subsection{Sparse Representation Learning}

Sparse representations based on PLM have garnered increasing interest due to their inherent advantages from lexical models. These approaches generally involve a term-importance component and/or a term-expansion mechanism. Various designs have been proposed to address these aspects. COIL \cite{coil} learns term-level dense representations for a contextualized lexical match, while uniCOIL \cite{unicoil} simplifies this by learning a single weight per term, extending methods like DeepCT \cite{deepct} and DeepImpact \cite{deepimpact}. SPLADE \cite{splade}, on the other hand, directly learns high-dimensional sparse representations capable of joint expansion and re-weighting through the Masked Language Modeling head of the PLM and sparse regularization. Recent Splade versions like \cite{hardnegatives} and \cite{spladev3} show the importance of mining stronger hard negatives and utilization of distillation losses to boost in-domain and out-of-domain retriever performance. There have also been works to combine both sparse and dense representations for better retrieval \cite{sparseembed} while maintaining linear time complexity.

\subsection{Echo embeddings}

Since LLMs are decoder-only causal language models, they are inherently weak at contextual representation due to unidirectional attention \cite{llm2vec}. To mitigate this \citet{echoembeddings} proposed to use repetition to overcome this bias. In this idea, they propose to use two occurrences of the same text which is passed to the language model, and take a mean representation of only the second occurrence. This way, all the tokens from the second occurrence can already be seen in the whole context, leading to better representations. We utilize echo embeddings instead of an expensive retraining of the model for bidirectional attention.

\section{Splade and Methodology}
In this section, we first describe in detail SPLADE \cite{splade}, alongside how to leverage diverse data without any hard negative mining and how to utilize causal LLMs for SPLADE training.

\subsection{SPLADE}
\textbf{Model.} SPLADE generates interpretable sparse representations by utilizing the BERT Wordpiece vocabulary for token expansion and the MLM head for assigning term weights. This way the model performs implicit vocabulary expansion without any need for providing manual expansion terms. The interpretability of SPLADE is maintained due to the tied weight matrices of the embedding and the MLM layer. For a query or document $t$, let $w_{i,j}$ denote the resulting importance of the $j$-th vocabulary token, for the input token $i$. Text representations are obtained by pooling such importance predictors over the input sequence, after a log-saturation effect. We thus consider by default the following formulation:
\[
w_j = \max_{\substack{i \in t}} \log \left( 1 + \text{ReLU}(w_{ij}) \right)
\]
and the ranking score $s(q, d)$ is given by the dot product between $q$ and $d$ representations.

\textbf{Training.} Given a query  $q$, a positive document  $d^+$, and additional in-batch negatives  $d^-_j$ (which are documents from other queries within the same batch), the model is trained through a combined approach. This involves optimizing a contrastive InfoNCE loss \cite{infonce}, as commonly employed in various prior works on learning first-stage rankers, alongside FLOPS regularization \cite{flops}. The regularization is directly applied to the representations in the vocabulary space to achieve the desired sparsity factor in the indices during inference:
\[
\mathcal{L} = \mathcal{L}_{\text{InfoNCE}} + \lambda_q \mathcal{L}^q_{\text{FLOPS}} + \lambda_d \mathcal{L}^d_{\text{FLOPS}}
\]

Since we rely only on in-batch negatives $d^-_j$, we reduce the cost of any additional hard negative mining as compared to previous SPLADE versions \cite{splade, spladev3, hardnegatives}.

\begin{table*}[]
\centering
\resizebox{2.0\columnwidth}{!}{%
\begin{tabular}{|c|ccccc|cc|}
\hline
\multirow{2}{*}{\textbf{Corpus}} &
  \multicolumn{5}{c|}{\textbf{Baselines}} &
  \multicolumn{2}{c|}{\textbf{Ours}} \\ \cline{2-8} 
 &
  \multicolumn{1}{c|}{\textbf{BM25}} &
  \multicolumn{1}{c|}{\textbf{ColBERTv2}} &
  \multicolumn{1}{c|}{\textbf{SPLADE++}} &
  \multicolumn{1}{c|}{\textbf{Spladev3}} &
  \textbf{ElserV2} &
  \multicolumn{1}{c|}{\textbf{\begin{tabular}[c]{@{}c@{}}Bert-base\\ +Sent-Trans\end{tabular}}} &
  \textbf{\begin{tabular}[c]{@{}c@{}}Echo-Mistral\\ +Sent-Trans\end{tabular}} \\ \hline
Dense/Sparse &
  \multicolumn{1}{c|}{Sparse} &
  \multicolumn{1}{c|}{Dense} &
  \multicolumn{1}{c|}{Sparse} &
  \multicolumn{1}{c|}{Sparse} &
  Sparse &
  \multicolumn{1}{c|}{Sparse} &
  Sparse \\ \hline
TREC-COVID &
  \multicolumn{1}{c|}{65.6} &
  \multicolumn{1}{c|}{73.8} &
  \multicolumn{1}{c|}{72.5} &
  \multicolumn{1}{c|}{74.8} &
  72.9 &
  \multicolumn{1}{c|}{63.7} &
  \textbf{76.79} \\ \hline
NFCorpus &
  \multicolumn{1}{c|}{32.5} &
  \multicolumn{1}{c|}{33.8} &
  \multicolumn{1}{c|}{34.5} &
  \multicolumn{1}{c|}{35.7} &
  36.7 &
  \multicolumn{1}{c|}{34.29} &
  \textbf{42.28} \\ \hline
NQ &
  \multicolumn{1}{c|}{32.9} &
  \multicolumn{1}{c|}{56.2} &
  \multicolumn{1}{c|}{53.3} &
  \multicolumn{1}{c|}{\textbf{58.61}} &
  55.9 &
  \multicolumn{1}{c|}{43.27} &
  55.96 \\ \hline
HotpotQA &
  \multicolumn{1}{c|}{60.3} &
  \multicolumn{1}{c|}{66.7} &
  \multicolumn{1}{c|}{69.3} &
  \multicolumn{1}{c|}{69.2} &
  67 &
  \multicolumn{1}{c|}{64.76} &
  \textbf{70.17} \\ \hline
FiQA-2018 &
  \multicolumn{1}{c|}{23.6} &
  \multicolumn{1}{c|}{35.6} &
  \multicolumn{1}{c|}{34.9} &
  \multicolumn{1}{c|}{37.4} &
  41.6 &
  \multicolumn{1}{c|}{39.86} &
  \textbf{57.71} \\ \hline
ArguAna &
  \multicolumn{1}{c|}{31.5} &
  \multicolumn{1}{c|}{46.3} &
  \multicolumn{1}{c|}{51.8} &
  \multicolumn{1}{c|}{50.96} &
  55.6 &
  \multicolumn{1}{c|}{51.56} &
  \textbf{56.21} \\ \hline
Touché-2020 &
  \multicolumn{1}{c|}{\textbf{36.7}} &
  \multicolumn{1}{c|}{26.3} &
  \multicolumn{1}{c|}{24.2} &
  \multicolumn{1}{c|}{29.29} &
  26.3 &
  \multicolumn{1}{c|}{19.44} &
  17.99 \\ \hline
Quora &
  \multicolumn{1}{c|}{78.9} &
  \multicolumn{1}{c|}{85.2} &
  \multicolumn{1}{c|}{84.9} &
  \multicolumn{1}{c|}{81.4} &
  84.69 &
  \multicolumn{1}{c|}{\textbf{87.05}} &
  86.68 \\ \hline
DBPedia &
  \multicolumn{1}{c|}{31.3} &
  \multicolumn{1}{c|}{44.6} &
  \multicolumn{1}{c|}{43.6} &
  \multicolumn{1}{c|}{\textbf{44.9}} &
  42.7 &
  \multicolumn{1}{c|}{35.32} &
  41.99 \\ \hline
SCIDOCS &
  \multicolumn{1}{c|}{15.8} &
  \multicolumn{1}{c|}{15.4} &
  \multicolumn{1}{c|}{16.1} &
  \multicolumn{1}{c|}{15.78} &
  16.2 &
  \multicolumn{1}{c|}{21.24} &
  \textbf{25.62} \\ \hline
FEVER &
  \multicolumn{1}{c|}{75.3} &
  \multicolumn{1}{c|}{78.5} &
  \multicolumn{1}{c|}{79.6} &
  \multicolumn{1}{c|}{79.6} &
  78.5 &
  \multicolumn{1}{c|}{70.64} &
  \textbf{84.43} \\ \hline
Climate-FEVER &
  \multicolumn{1}{c|}{21.3} &
  \multicolumn{1}{c|}{17.6} &
  \multicolumn{1}{c|}{\textbf{23.7}} &
  \multicolumn{1}{c|}{23.3} &
  27.1 &
  \multicolumn{1}{c|}{22.11} &
  22.83 \\ \hline
SciFact &
  \multicolumn{1}{c|}{66.5} &
  \multicolumn{1}{c|}{69.3} &
  \multicolumn{1}{c|}{71} &
  \multicolumn{1}{c|}{70.96} &
  71.8 &
  \multicolumn{1}{c|}{69.31} &
  \textbf{77.24} \\ \hline
\textbf{Avg} &
  \multicolumn{1}{c|}{44.02} &
  \multicolumn{1}{c|}{49.95} &
  \multicolumn{1}{c|}{50.72} &
  \multicolumn{1}{c|}{51.68} &
  52.07 &
  \multicolumn{1}{c|}{47.89} &
  \textbf{55.07} \\ \hline
\end{tabular}
}
\caption{Zero shot ndcg@10 $\uparrow$ on BEIR (13). For comparison, we report results directly from corresponding papers, where the evaluation is generally done on the subset of 13 readily available BEIR datasets.}
\label{tab:results}
\end{table*}

\subsection{Scaling SPLADE for Causal decoder LLMs}

Learned Sparse Retriever models maintain their interpretability by expanding the input sequence into a set of related tokens, which leads to better retrieval. This interpretability is ensured by the tied weight matrices of the embedding layer and the language modeling head \cite{weight-tying}. Models like BERT \cite{devlin-etal-2019-bert} are pre-trained with tied weight matrices, and thus no intervention is required during SPLADE training apart from keeping them tied. Untying the weights would lead the language modeling layer to be equivalent to a linear layer, which does not correspond to any token for every logit. Many LLMs today are trained with untied weight matrices \cite{llama2, mistral}, which is bad for SPLADE training since the latent space is not bound to generate interpretable logits. Freezing the embedding and the output layer can still lead to uninterpretable logits and defeat the purpose of sparse retrieval systems. Another approach could be to reinitialize the language modeling head layer with the embedding layer and then retrain on a causal LM objective with tied weights. However, this is expensive and can also lead to loss of information. We thus use LoRA finetuning \cite{lora} to avoid this altogether. This way, we also reduce the number of parameters to be trained while maintaining interpretability. Another problem of causal LLMs is the weak representation of past context due to unidirectional attention. We utilize Echo embeddings \cite{echoembeddings} to improve the representation of the LLM. This effectively doubles the sequence length during training and inference. We refer to this model as Echo-Mistral SPLADE in our experiments.

\section{Experiments and Results}

In this section, we discuss our experimental setup for all the proposed methods and discuss their potential advantages and limitations through empirical results.

\subsection{Training}
All the reference models are trained on the MS MARCO passage ranking dataset, which contains 8.8M passages and 500k training queries with shallow annotations. These models have been trained in multiple stages with each stage introducing more and more hard negatives and better margin scores for distillation. Instead, we rely on the same training mechanism with much more diverse training data. Specifically, we use the Sentence-transformers embedding data (15.5M subset) which has been previously used for sentence similarity tasks. Instead of mining hard negatives and calculating distillation scores which requires a lot of excess compute and effort. We rely only on in-batch negatives for training, with a batch size of 512. We use the same number of steps (150k) as other splade variants with echo embeddings during training and inference for our Echo-Mistral SPLADE model.

We initialized the smaller SPLADE models with the BERT-base checkpoint and the LLM with Mistralv3\footnote{\url{https://huggingface.co/mistralai/Mistral-7B-v0.3}}. Models are trained with the ADAM optimizer, using a learning rate of 2e-5 with linear scheduling and a warmup of 6000 steps. We consider a maximum length of 256 for input sequences. To mitigate the contribution of the regularizer at the early stages of training, we follow \cite{flops} and use a scheduler for \(\lambda\), quadratically increasing \(\lambda\) at each training iteration, until a given step (50k in our case), from which it remains constant. For training the larger Echo-Mistral-7B SPLADE model we use QLoRA finetuning \cite{qlora} with $\alpha=8$, $rank=16$, and a dropout value of 0.1 over the projection layers. Models are trained using PyTorch \cite{pytorch} and HuggingFace transformers \cite{huggingface}, on 4 NVIDIA DGX A100 GPUs with 80GB memory.

\subsection{Sentence Transformers embedding data}

Sparse retrievers have been trained only on individual datasets like the MSMarco passage ranking dataset which has 8.8M passages and 500k training queries. MSMarco is a high-quality retrieval dataset that is generated from Bing search logs. Specifically, it contains anonymized data derived from real user queries submitted to the Bing search engine. Instead of MSMarco, we rely on a large-scale sentence embedding dataset called Sentence-Transformer\footnote{\url{https://huggingface.co/collections/sentence-transformers/embedding-model-datasets-6644d7a3673a511914aa7552}} that has shown improvements over BEIR benchmark for many DPR models \cite{sbert, senttrans1, senttrans2, senttrans3, senttrans4, llm2vec}. Specifically, we take the 15.5M sample subset and set their sampling probabilities proportional to their dataset size. We list all the dataset names in Appendix \ref{app: dataset}. We use this dataset for Echo-Mistral-SPLADE training without any hard negative mining or distillation.

\subsection{Evaluation}
We report nDCG@10 for comparing zero-shot performance of our models on the BEIR benchmark \cite{beir}. We rely on the subset of 13 readily available datasets to compare with other baselines, thus we do not consider CQADupstack, BioASQ, Signal-1M, TREC-NEWS, and Robust04 BEIR datasets for evaluation.

\begin{table}[]
\centering
\resizebox{0.99\columnwidth}{!}{%
\begin{tabular}{|c|c|c|c|}
\hline
\textbf{Corpus} &
  \textbf{\begin{tabular}[c]{@{}c@{}}LLM2Vec\\ Mistral\\ Supervised\end{tabular}} &
  \textbf{\begin{tabular}[c]{@{}c@{}}Echo Embeddings\\ Mistral\end{tabular}} &
  \textbf{\begin{tabular}[c]{@{}c@{}}Echo-Mistral\\ SPLADE\end{tabular}} \\ \hline
Dense/Sparse  & Dense          & Dense          & Sparse         \\ \hline
TREC-COVID    & \textbf{77.69} & 76.02          & 76.79          \\ \hline
NFCorpus      & 39.33          & 37.18          & \textbf{42.28} \\ \hline
NQ            & \textbf{61.7}  & 61.51          & 55.96          \\ \hline
HotpotQA      & 74.07          & \textbf{75.75} & 70.17          \\ \hline
FiQA-2018     & 53.11          & 54.11          & \textbf{57.71} \\ \hline
ArguAna       & \textbf{57.48} & 56.52          & 56.21          \\ \hline
Touché-2020   & \textbf{22.18} & 18.92          & 17.99          \\ \hline
Quora         & 87.75          & \textbf{89.33} & 86.68          \\ \hline
DBPedia       & \textbf{49.58} & 48.19          & 41.99          \\ \hline
SCIDOCS       & 22.5           & 17.73          & \textbf{25.62} \\ \hline
FEVER         & 89.4           & \textbf{91.14} & 84.43          \\ \hline
Climate-FEVER & 35.19          & \textbf{37.07} & 22.83          \\ \hline
SciFact       & \textbf{78.86} & 73.57          & 77.24          \\ \hline
\textbf{Avg}  & \textbf{57.6}  & 56.69          & 55.07          \\ \hline
\end{tabular}
}
\caption{Comparison with DPR models trained on Mistral-7B on zero-shot nDCG@10 $\uparrow$ on BEIR (13). The training sets of NQ, FEVER, HotpotQA, and Climate-FEVER were included while training LLM2Vec \cite{llm2vec} and Mistral with Echo Embeddings \cite{echoembeddings}, so the comparison with our sparse model is not entirely fair.}
\label{tab:my-table}
\end{table}


\subsection{Results and Discussion}

We compare our model with the most prelevant learned sparse models from SPLADE family and its subsequent variations and extensions like Elserv2 \cite{hardnegatives, splade, spladev3, elastic_elser_2024} and also with ColBERT\cite{colbert} which is a state-of-the-art dense baseline. 
Our results can be summarized in Table \ref{tab:results}, we can see that scaling with echo embeddings work significantly well over the same dataset, achieving state-of-the-art results for sparse retrieval on BEIR-13 without any need for complicated training and hard negative mining. 

For a fair comparison with similar-sized LLM-based dense retriever models, we also compare our results with LLM2Vec on Mistral7B in a supervised setting after MNTP and SimCSE pretraining \cite{llm2vec} and Echo Embeddings on Mistral7B with pooled representation of the second occurrence \cite{echoembeddings}. Firstly, we do not use any of the BEIR train sets while training our LSR model so all the numbers reported are purely zero shot as compared to LLM2Vec and Echo Embeddings which use train sets of NQ, Fever, and HotpotQA during their training. In Table \ref{tab:my-table} we can see that a similar-sized Echo-Mistral SPLADE model achieves competitive performance to dense models. If we exclude in-domain performance for all models, we see Echo-Mistral SPLADE surpassing the performance of Echo Embeddings but lagging behind LLM2Vec which shows performance gains due to unsupervised pretraining. This is a strong indicator of how well LSR systems can perform in practice.

\section{Conclusion}

In this paper, we present Echo-Mistral SPLADE, which uses Mistral-7B LLM with Echo embeddings as the backbone for learned sparse retrieval training. Echo-Mistral-SPLADE achieves state-of-the-art performance on BEIR benchmark outperforming existing sparse retrieval models including SPLADE and its variants. We also compare with a similar-sized dense model and show that Echo-Mistral-SPLADE can outperform dense models in out-of-domain setting, which can be further improved with unsupervised pretraining. We believe such diverse data and the model scale is the leading reason for improvement but further exploration like joint distillation and hard negatives can further improve this performance. With this, we aim to motivate the use of LSR systems in practice due to their low inference latency and interpretability over dense retriever systems.

\bibliography{anthology,custom}

\appendix

\section{Dataset Details}
\label{app: dataset}
In this section we list down the dataset details and sampling probabilities during training. The probabilities are calculated by normalizing the size of each dataset and we exclude any dataset that conflicts with our evaluation on zero-shot out-of-domain BEIR retrieval datasets \cite{beir}.

\begin{table}[h!]
\centering
\resizebox{0.95\columnwidth}{!}{%
\begin{tabular}{|c|c|}
\hline
\textbf{Dataset Name} & \textbf{Sampling ratio} \\ \hline
gooaq\_pairs & 20.53 \\ \hline
yahoo\_answers\_title\_answer & 8.17 \\ \hline
msmarco\_triplets & 3.43 \\ \hline
stackexchange\_title\_title & 2.08 \\ \hline
eli5\_question\_answer & 2.22 \\ \hline
yahoo\_answers\_title\_question & 4.50 \\ \hline
squad\_pairs & 0.60 \\ \hline
yahoo\_answers\_question\_answer & 4.64 \\ \hline
wikihow & 0.88 \\ \hline
amazon\_qa & 17.08 \\ \hline
quora question pairs & 1.02 \\ \hline
stackexchange\_title\_body\_title\_body & 1.71 \\ \hline
stackexchange\_body\_body & 1.71 \\ \hline
agnews & 7.89 \\ \hline
AllNLI & 2.14 \\ \hline
npr & 4.05 \\ \hline
specter\_train\_triples & 2.59 \\ \hline
SimpleWiki & 0.70 \\ \hline
altlex & 0.77 \\ \hline
ccnews\_title\_text & 4.19 \\ \hline
sentence\_compression & 1.23 \\ \hline
TriviaQA\_pairs & 0.50 \\ \hline
cnn\_dailymail & 1.96 \\ \hline
flickr30k\_captions & 1.08 \\ \hline
xsum & 1.54 \\ \hline
coco\_captions & 2.82 \\ \hline
\end{tabular}%
}
\caption{Dataset Sampling Probabilities}
\label{tab:sampling_probabilities}
\end{table}

\end{document}